\documentclass{article}

\usepackage{graphicx}
\usepackage{latexsym}
\usepackage[fleqn]{amsmath}
\usepackage{amssymb}
\usepackage{amsbsy}
\usepackage{subcaption}

\newtheorem{remark}{Remark}

\begin{document}

\title{Collision Avoidance Control for a Two-wheeled Vehicle under Stochastic Vibration using an Almost Sure Control Barrier Function}

\author{Taichi Arimura\footnotemark[1], Y\^uki Nishimura\footnotemark[2], Taichi Ikezaki\footnotemark[2] and Daisuke Tabuchi\footnotemark[1]}
\date{March 30, 2026\footnotemark[3]}

\renewcommand{\thefootnote}{\fnsymbol{footnote}}
\footnotetext[1]{Kagoshima University}
\footnotetext[2]{Okayama University}
\footnotetext[3]{This work has been submitted to the SICE Journal of Control, Measurement, Systems and Integration for possible publication. Copyright may be transferred without notice, after which this version may no longer be accessible.}

\renewcommand{\thefootnote}{\arabic{footnote}}
\maketitle

\begin{abstract}
In recent years, many control problems of autonomous mobile robots have been developed. In particular, the robots are required to be safe; that is, they need to be controlled to avoid colliding with people or objects while traveling. In addition, since safety should be ensured even under irregular disturbances, the control for safety is required to be effective for stochastic systems. In this study, we design an almost sure safety-critical control law, which ensures safety with probability one, for a two-wheeled vehicle based on the stochastic control barrier function approach. In the procedure, we also consider a system model using the relative distance measured by a 2D LiDAR. The validity of the proposed control scheme is confirmed by experiments of a collision avoidance problem for a two-wheeled vehicle under vibration.
\end{abstract}

\section{Introduction}
The development of autonomous mobile robots, such as cleaning robots and automatic guided vehicles, is promoted to improve work efficiency and solve labor shortages \cite{takeuchi,maekawa}. Because autonomous mobile robots are often used in residential spaces and factories with surrounding obstacles, they are required to be safe enough to avoid collisions with people or objects while traveling. 

The collision avoidance problems are solved via various methods: the artificial potential field method by Rimon and Koditschek \cite{artificial_potential_fuction}, the dynamic window approach by Fox, Burgard and Thrun\cite{DWA} and Hahn\cite{DWA2}, the nearness diagram method by Minguez and Montano \cite{ND}, a navigation method considering moving obstacles by Tsubouchi et al. \cite{planning_and_navigation}, a collision avoidance method based on fuzzy inference by Maeda and Takegaki \cite{fuzzy_inference}, and a navigation method considering the future behavior of dynamic obstacles by Z.~Zhang et al. \cite{one_shot}.

Recently, a safety-critical control approach based on control barrier functions \cite{Ames_Coogan_} is attracting attention as a scheme for maintaining the safety of mobile robots because of its simple and strong control design. Kimura and Nishimoto \cite{kimurakai} propose a design method for collision avoidance control of a two-wheeled vehicle using measured data of the distance between the obstacle and the vehicle via 2D LiDAR based on \cite{kimuraiecon,nakamura}. 

At the same time, the actual traveling environment of mobile robots is often influenced by irregular disturbances. Therefore, designing a control law that achieves the safety-critical control objective even under the influence of irregular disturbances is required. More clearly, the safety is preferable to hold with 100\%; that is, {\it almost surely}, even when white noise vibrates the robots. Nishimura and Hoshino \cite{nishimura_hoshino_RCBF} propose a design procedure for an {\it almost sure safety-critical control law} for nonlinear stochastic systems including Gaussian white noise. Therefore, the combination of the methods in \cite{kimurakai} and \cite{nishimura_hoshino_RCBF} is effective for controlling a two-wheeled vehicle in the real-world environment.

In this paper, we apply the almost sure safety-critical control scheme by Nishimura and Hoshino \cite{nishimura_hoshino_RCBF} to the collision avoidance system of a two-wheeled vehicle by Kimura and Nishimoto \cite{kimurakai}, and then derive a control strategy that achieves collision avoidance with 100\% against the existence of irregular disturbances.

The organization of this paper is as follows. 
In Section~\ref{sec:preliminary}, we briefly summarize the previous works used in this study. 
In Section~\ref{control_target}, we present the stochastic system of the two-wheeled vehicle considered as the control target. 
In Section~\ref{control_design}, we describe the almost sure safety control law for avoiding collisions between the vehicle and obstacles. 
In Section~\ref{experiment}, we confirm and discuss the validity of the proposed collision avoidance controller via simulations and experiments.  
Finally, in Section~\ref{conclusion}, we conclude this paper.

{\it Notation. }
We introduce the notation used throughout this paper. $\mathbb{R}^n$ denotes the $n$-dimensional Euclidean space and especially $\mathbb{R}:=\mathbb{R}^1$. $\mathbb{R}_{\geq 0}$ also denotes the set of nonnegative real numbers. 
The Lie derivative of a smooth mapping $W:\mathbb{R}^n\rightarrow\mathbb{R}$ along a mapping $F=(F_1,\ldots,F_q):\mathbb{R}^n\rightarrow\mathbb{R}^{n\times q}$ is defined as
\begin{align}
    L_FW(x)=\left(\frac{\partial W}{\partial x}F_1(x),\ldots,\frac{\partial W}{\partial x}F_q(x)\right).
\end{align}
The boundary of a set $\chi$ is denoted by $\partial \chi$. The mapping $w : [0,\infty)\rightarrow\mathbb{R}$ denotes a one-dimensional standard Wiener process. The differential form of the It\^o integral of a mapping $\sigma:\mathbb{R}^n\rightarrow\mathbb{R}$ along $w$ is denoted by $\sigma(x)dw$. The trace of a square matrix $\mathrm{A}$ is denoted by $\mathrm{tr}[\mathrm{A}]$.

\begin{figure}[t]
    \begin{minipage}[t]{0.48\columnwidth}
        \centering
        \includegraphics[width=.9 \linewidth]{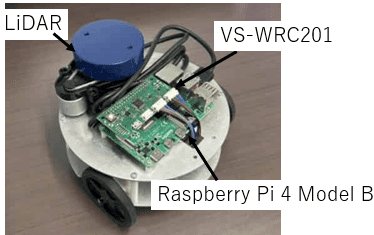}
        \caption{Lightrover.}
        \label{fig:lightrover}
    \end{minipage}
    \hspace{0.04\columnwidth}        
    \begin{minipage}[t]{0.48\columnwidth}
        \centering
        \includegraphics[width=.9 \linewidth]{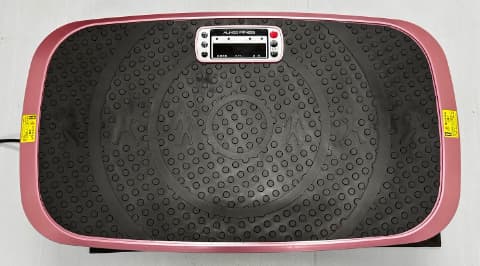}
        \caption{Balancewave rose FAV4318P.}
        \label{fig:balancewave}
    \end{minipage}
\end{figure}

{\it Experimental Environment. }
We use a Lightrover made by Vstone Co., Ltd., shown in Fig.~\ref{fig:lightrover} as a two-wheeled vehicle, a YDLiDAR X2 made by Shenzhen EAI Technology Co., Ltd, as a 2D LiDAR, and a Balancewave Rose FAV4318P made by ALINCO Inc., shown in Fig.~\ref{fig:balancewave} as a noise source, which is capable of applying mixed vibration combining three-dimensional vibration and micro vibration.

\section{Preliminary: Safety-critical Control}\label{sec:preliminary}
\subsection{Safety-critical Control}\label{subsec:cbf}
In this subsection, we briefly summarize a safety-critical control law proposed in \cite{nakamura}. Consider an input-affine nonlinear control system
\begin{align}
    \dot{x}(t)=f(x(t))+g(x(t))(u_o(t)+u(t)),
    \label{nakamura_system}
\end{align}
where $x: [0,\infty) \to \mathbb{R}^{n}$ is a state, $u_o: [0,\infty) \to \mathbb{R}^m$ is a preinput assumed to be continuous, $u: [0,\infty) \to \mathbb{R}^m$ is a compensator for safety-critical control, $f:\mathbb{R}^n \to \mathbb{R}^n$ and $g:\mathbb{R}^n \to \mathbb{R}^{n\times m}$ are both assumed to be locally Lipschitz continuous, and an initial state is given as $x_0 = x(0) \in \mathbb{R}^n$. 

Consider an open set $\chi \subset \mathbb{R}^n$ and a continuously differentiable function $B:\chi \rightarrow \mathbb{R}$. If the following assumptions:
\begin{description}
    \item[(A1)] $B(x)\geq0$ for all $x\in\chi$,
    \item[(A2)] for any $L \geq 0$, $\{x \in \chi \mid B(x)\leq L \}$ is compact,
    \item[(A3d)] for any continuous mapping $u_o:\mathbb{R}\rightarrow\mathbb{R}^m$, there exist nonnegative constants $C,K\geq0$ such that
    \begin{align}
        \inf_{u\in \mathbb{R}^m}\dot{B}(x,u_o,u)<KB(x)+C,
    \end{align}
\end{description}
are all satisfied, then $\chi$ and $B(x)$ are said to be a {\it safe set} and a {\it control barrier function (CBF)}, respectively. 
Under the assumptions (A1), (A2) and (A3d), designing $u=\phi_D(t,x)$ with
\begin{align} \label{eq:det-ctrl}
    \phi_D(t,x)\!=\!\left\{
        \begin{array}{ll}
            \!\!-\frac{I_d(x,u_o)-J_d(x)}{||L_gB(x)||^2}\!(L_g B(x))^T\!\!, & \!I_d(x,u_o)\!>\!J_d(x), \\
            0, & \mathrm{otherwise},
        \end{array}\right.
\end{align}
where 
\begin{align}
    &I_d(x,u_o)=L_fB(x)+L_gB(x)u_o,\\
    &J_d(x)=KB(x)+C,
\end{align}
the system \eqref{nakamura_system} is safe in $\chi$; that is, the trajectory $x$ keeps staying in $\chi$ for any initial value $x_0 \in \chi$. 

\subsection{Almost Sure Safety-critical Control}\label{Nishimura_Hoshino} 
In this subsection, we consider an almost sure safety-critical control law that keeps the state in the safe set with probability one, proposed in \cite{nishimura_hoshino_RCBF}.

Assuming that the system \eqref{nakamura_system} is influenced by Gaussian white noise, we obtain a stochastic system
\begin{align}
    dx(t)=&\{f(x(t))+g(x(t))(u_o(t)+u(t))\}dt 
        +\sigma(x(t))dw(t),
    \label{stochastic_system}
\end{align}
where $f,g,u_o,u,x$ and $x_0$ are the same as those in \eqref{nakamura_system}, and $\sigma:\mathbb{R}^n \rightarrow \mathbb{R}^{n \times d}$ is assumed to be locally Lipschitz continuous. 
Let
\begin{align}
    &L_\sigma^I(y(x)):=\frac{1}{2}\mathrm{tr}\left[\sigma(x)\sigma(x)^T
    \left[\frac{\partial}{\partial x}\left[\frac{\partial y}{\partial x}\right]^T\right](x)\right],\\
    &\mathcal{L}_{f,g,\sigma}(u,u_o,y(x)):= 
    (L_fy)(x)+(L_gy)(x)(u+u_o)+L_\sigma^I(y(x)), 
\end{align}
where $B:\chi \rightarrow \mathbb{R}$ is twice continuously differentiable, and an open set $\chi \subset \mathbb{R}^n$. If (A1), (A2) and
\begin{description}
    \item[(A3s)] for any continuous mapping $u_o:\mathbb{R}\rightarrow\mathbb{R}^m$, there exist nonnegative constants $\gamma \geq 0$ such that
    \begin{align}
        \inf_{u\in \mathbb{R}^m} \mathcal{L}_{f,g,\sigma}(u,u_o,B(x))\leq \gamma B(x),
    \end{align}
\end{description}
are all satisfied, then $\chi$ and $B(x)$ are said to be a {\it safe set} and a {\it almost sure reciprocal control barrier function (AS-RCBF)}, respectively. 

Under the assumptions (A1), (A2) and (A3s), design 
    \begin{align}
    &\phi_N(t,x)=\left\{
        \begin{array}{ll}
            \psi(t,x), & I>J\cap L_gB\neq0, \\
            0, & I\leq J \cup L_gB=0,
        \end{array}\right.
    \label{phi_N}
    \end{align}
    where
    \begin{align}
        &I(u_o,B(x)):=\mathcal{L}_{f,g,\sigma}(0,u_o,B(x)),
        \label{I(B)} \\
        &J(B(x)):=\gamma B(x),
        \label{J(B)} \\
        &\psi(t,x) := -\frac{I(u_o,B(x))-J(B(x))}{L_gB(x)L_gB(x)^T}L_gB(x)^T.  \label{eq:psi}
    \end{align}
    If
    \begin{align}
        L_f(B(x))+L^I_\sigma(B(x)) > \gamma B(x)
    \end{align}
    holds on $L_gh=0$, then the compensator $u=\phi_N(t,x)$ is continuous and makes the system \eqref{stochastic_system} safe in $\chi$ with probability one; that is, the trajectory $x$ keeps staying in $\chi$ for any initial value $x_0 \in \chi$ with 100\% probability \cite{nishimura_hoshino_RCBF}. In addition, if $\sigma = 0$ for all $x$, $\phi_N = \phi_D$ with $K=\gamma$ and $C=0$. 

\begin{remark}
     In this paper, the preinput $u_o$ is assumed to be time varying; that is, $u_o(t)$, while it is assumed to be time invariant; that is, $u_o(x(t))$ in \cite{nishimura_hoshino_RCBF}. In fact, the results from \cite{nishimura_hoshino_RCBF} are directly applied to $u_o(t)$ because the results are based on the forward invariance in probability (FIiP), a sufficient condition for which is given in \cite{nishimura2018}. This remark is also stated in \cite{henmi2026}, and the same result on deterministic systems is described in \cite{nakamura} as shown in Subsection~\ref{subsec:cbf}.
\end{remark}

\section{Target System}\label{control_target}

\subsection{Dynamics for a Vehicle with 2D LiDAR}\label{system_model}

\begin{figure}[t]
    \begin{minipage}[t]{0.48\columnwidth}
        \centering
        \includegraphics[width=.9 \linewidth]{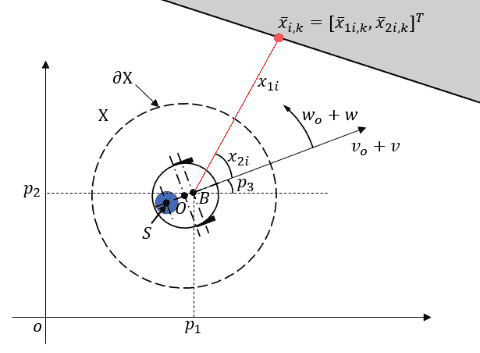}
        \caption{A system model of a two-wheeled vehicle.} 
        \label{fig:lightrover_model}
    \end{minipage}
    \hspace{0.04\columnwidth}        
    \begin{minipage}[t]{0.48\columnwidth}
        \centering
        \includegraphics[width=0.8 \linewidth]{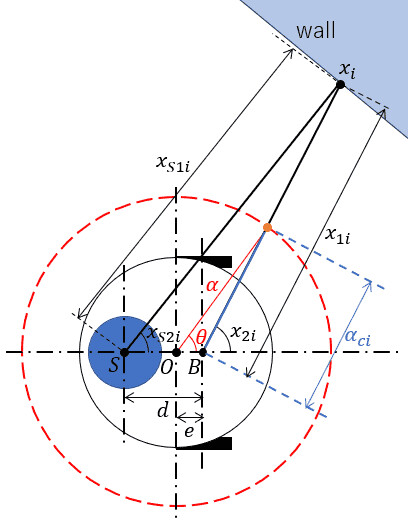}
        \caption{Relationships among constants and variables.}
        \label{fig:sensor_pos}
    \end{minipage}
\end{figure}

In this subsection, we consider the dynamics of the two-wheeled vehicle, 
shown in Fig.~\ref{fig:lightrover_model}, 
based on \cite{kimurakai}; the difference between our vehicle and the vehicle in \cite{kimurakai} will be stated in Remark~\ref{rem:differ} at the end of this section.

We define the translational velocity $v_o+v$ and the rotational velocity $w_o+w$ as inputs, respectively, where $v_o$ and $w_o$ are preinputs (previously given inputs), and $v$ and $w$ are compensators for safety-critical control. Adding the inputs for the vehicle, we obtain
\begin{align}
    \dot{p}_1&=(v_o+v)\cos{p_3}, \label{kimurakai1}\\
    \dot{p}_2&=(v_o+v)\sin{p_3}, \label{kimurakai2}\\
    \dot{p}_3&=(w_o+w),         \label{kimurakai3}
\end{align}
where $p_1$ and $p_2$ describes the center position $B$ of the axle, and $p_3$ as the angular difference from the $p_1$ axis. 

We obtain the distance from the vehicle to an obstacle using the 2D LiDAR mounted on the vehicle, where the forward direction is defined as $0$~rad, the measurement range is $2\pi$~rad, and the number of measurement points is denoted by $N$. 
For $i\in\{1,\ldots,N\}$, the distance and the angle of the $i$-th point are denoted by $x_{1i}\in\mathbb{R}_{\geq0}$ and $x_{2i}\in[-\pi,\pi)$, respectively. 

Set the time sequence for measuring as $t_0, t_1, \ldots, t_k$ with $t_0=0$, the time intervals as $[t_k,t_{k+1})$ for any $k = 1,2,\ldots$, and assume that $t_{k+1} - t_k$ is a constant for any $k=1,2,\ldots$. The absolute coordinates of the $i$-th point observed at time $t=t_k$ with $k \in \{ 1,2,\ldots \}$ are represented by 
\begin{align}
    \bar{x}_{i,k} = \begin{bmatrix} \bar{x}_{1i,k} \\ \bar{x}_{2i,k} \end{bmatrix} = \begin{bmatrix}
        x_{1i} \cos (x_{2i} + p_3) + p_1 \\ x_{1i} \sin (x_{2i} + p_3) + p_2
    \end{bmatrix}. \label{eq:xik}
\end{align}
Assuming that the obstacle is time-invariant and the vehicle acts sufficiently slowly, $\bar{x}_{i,k}$ is constant. Hence, differentiating \eqref{eq:xik}, using $\dot{\bar{x}}_{1,k}=\dot{\bar{x}}_{2,k}=0$ and defining $x_i = [x_{1i},x_{2i}]^T$, we obtain
\begin{align}
    \dot{x}_i= g_i(x)
        \begin{bmatrix}
            v_o+v \\
            w_o+w
        \end{bmatrix},\quad
    g_i(x)=
    \begin{bmatrix}
        -\cos{x_{2i}} & 0 \\
        \frac{\sin{x_{2i}}}{x_{1i}} & -1 
    \end{bmatrix}, 
    \label{kimurakai6}
\end{align}
for $i=1,\ldots,N$ in $t \in [t_k,t_{k+1})$. 
Since we will consider $t \in [t_k,t_{k+1})$ hereafter, we omit the subscript $k$ for simplicity notation, as in \cite{kimurakai}.

Setting 
 \begin{align}
     &x =
     \begin{bmatrix}
         x_1^T, \dots, x_i^T, \dots, x_N^T 
     \end{bmatrix}^T, 
    \quad g(x)=
    \begin{bmatrix}
        g_1^T(x_1) & \ldots &
        g_i^T(x_i) & \ldots &
        g_N^T(x_N)
    \end{bmatrix}^T,
    \label{x_g(x)}
\end{align}
we obtain the following system model: 
\begin{align}
    \dot{x}=g(x)(u_o+u),
    \label{point_cloud_system_kimura}
\end{align}
where $x \in \mathbb{R}^n$ with $n=2N$. 

\subsection{Distance from the Sensor to the Axle}

In this subsection, we derive the distance and the angle from the center of the driving wheels based on the measurement results by the sensor. In our vehicle shown in Fig.~\ref{fig:sensor_pos}, the center of the sensor $S$ is not equivalent to the center of the driving wheels $B$ because of the specifications. This implies that we have to clarify the relationship between $S$ and $B$. 

Let $d$ be the distance between $S$ and $B$. For $i \in \{1,2,\ldots,N\}$, $x_{S1i}$ and $x_{S2i}$ are the distance and the angle between the measured point $x_i$ and the sensor, respectively. 
Then, considering $S$ as the origin, $x_i$ is expressed as 
$(x_{S1i}\cos{x_{S2i}},x_{S1i}\sin{x_{S2i}})$. 
Therefore, $x_{1i}$ and $x_{2i}$ are given by
\begin{align}
    x_{1i}&=\sqrt{(x_{S1i}\cos{x_{S2i}}-d)^2+(x_{S1i}\sin{x_{S2i}})^2},\\
    x_{2i}&=\arccos\frac{x_{S1i}\cos{x_{S2i}}-d}{x_{1i}},
\end{align}
respectively. 

\subsection{Allowable Distance from the Axle to an Obstacle}\label{collision_judgment_distance}

In this subsection, we consider the allowable distance to an obstacle. 
We set our control objective to keep the distance between $O$ and an object within the desired value $\alpha>0$. Therefore, we consider the threshold $\alpha_{ci}$ of the distance between $B$ and an obstacle. 

Denoting $e$ as the distance between $O$ and $B$, 
\begin{align}
    &\alpha_{ci}=\sqrt{(\alpha\cos{\theta}-e)^2+(\alpha\sin{\theta})^2},\\
    &\tan{x_{2i}} = \frac{\alpha \sin \theta}{\alpha \cos \theta - e},\\
    &\alpha \sin \theta = \alpha_{ci} \sin x_{2i}.
\end{align}
are obtained. Thus, assuming $\alpha > e > 0$, $\alpha_{ci}$ is represented as the function of $x_{2i}$:
\begin{align}\label{eq:alpha_ci}
    \alpha_{ci} = -e \cos x_{2i} + \sqrt{\alpha^2 -e^2 \sin^2 x_{2i}}.
\end{align}

\subsection{Stochastic System Model}

In this subsection, we consider the situation where noise is added to our system model.

Letting $\sigma_i=[c_1,c_2]^T$ and assuming that a Gaussian white noise is added into the system \eqref{kimurakai6}, we obtain a stochastic system
\begin{align}
    dx_i=g_i(x_i)(u_o+u)dt+\sigma_i dw
    \label{point_system_noise}
\end{align}
for each $i \in \{1,2,\ldots,N\}$. Then, defining
\begin{align}
    \sigma=
    \begin{bmatrix}
        \sigma_1^T, \dots, \sigma_i^T, \dots, \sigma_N^T
    \end{bmatrix}^T,
\end{align}
we obtain our target system
\begin{align}
    dx=g(x)(u_o+u)dt+\sigma dw,
    \label{lightrover_system}
\end{align}
which is a stochastic version of \eqref{point_cloud_system_kimura}. 

\begin{remark}\label{rem:differ}
Our vehicle model in \eqref{lightrover_system} has differences from the vehicle in \cite{kimurakai}. First, in our vehicle, the center $B$ of the driving wheels is not equivalent to the sensor position $S$. Second, our vehicle is circular. Third, we consider the angle $x_{2i}$ as a state variable of the system model to ensure consistency with the theory. And finally, we assume the system is vibrated by stochastic noise.
\end{remark}

\section{Almost Sure Safety-critical Control Law}\label{control_design}
In this section, we propose an AS-RCBF for the system~\eqref{lightrover_system} and design an almost sure safety control law to avoid collisions between the vehicle and obstacles based on the design procedure of a CBF in \cite{kimurakai} and the almost sure safety-critical control theory in Section~\ref{Nishimura_Hoshino}.

For the $i$-th point $x_i$, we define 
\begin{align}
    &\chi_i=\{x_i \mid x_{1i}>\alpha_{ci}\},    \label{chi_i_kimura} \\
    &B_i(x_i)=\frac{1}{x_{1i}-\alpha_{ci}}.  
\end{align}
Because our goal is to keep $x_{1i} > \alpha_{ci}$ for all $i$, we define
\begin{align}
    &\chi=\prod_{i=1}^{N}\chi_i,   \label{chi_k} \\
    &B(x)=\sum_{i=1}^{N}B_i(x_i),    \label{B(x)}
\end{align}
as a safe set and an AS-RCBF for \eqref{lightrover_system}, respectively.


Then, $\psi(x)$ in \eqref{eq:psi}, which is the part of the compensator $\phi_N$, results in 
\begin{align}
    \psi(t,x)=-\begin{bmatrix} v_o \\ w_o \end{bmatrix}+ \frac{\gamma B(x) - L^I_\sigma(B(x))}{||L_gB(x)||^2} (L_g B(x))^T,
\label{cal_phiN}
\end{align}
where
\begin{align}
    L_gB(x) &= \sum_{i=1}^N L_{g_i}B_i 
         = \sum_{i=1}^N \frac{1}{(x_{1i}-\alpha_{ci})^2} \begin{bmatrix}
        \cos x_{2i} + \frac{\sin{x_{2i}}}{x_1i} \alpha'_{ci} \\ 
        -\alpha'_{ci}
    \end{bmatrix}^T, \label{eq:lgb} \\
    L^I_\sigma(x) &= \frac12 \sum_{i=1}^N \sigma^T_i \frac{\partial}{\partial x_i} \left[ \frac{\partial B_i}{\partial x_i} \right]^T \sigma_i 
        = \sum_{i=1}^N \frac{1}{(x_{1i}-\alpha_{ci})^3} \sigma^T_i \begin{bmatrix}
            1 & - \alpha'_{ci} \\ -\alpha'_{ci} & \beta_i
        \end{bmatrix} \sigma_i,\\
    \alpha'_{ci} &:=\frac{\partial \alpha_{ci}}{\partial x_{2i}} 
        = e \sin{x_{2i}} - \frac{e^2 \sin{(x_{2i})} \cos{(x_{2i}})}{\sqrt{\alpha^2-e^2 \sin^2 x_2}}, \\
    \beta_i &:= \alpha'_{ci} \left( \alpha'_{ci} + \frac12 (x_{1i}-\alpha_{ci}) \frac{\partial \alpha'_{ci}}{\partial x_{2i}} \right).
\end{align}

\begin{remark}
    While $B(x)$ in \eqref{B(x)} does not satisfy (A2), the condition is merely a theoretical requirement; that is, if (A2) is satisfied, the existence of a global solution in time is ensured. We can design a theoretically accurate AS-RCBF as
    \begin{align}
        \bar{B}(x) = B(x) + \varepsilon P(x),
    \end{align}
    where $P:\chi \to [0,\infty)$ is a design function such that it is compact in $\{ x \in \chi | P(x) \le L\}$ for any $L>0$, and $\varepsilon > 0$ is a design parameter. Because we can choose $\varepsilon$ to be arbitrarily small, we ignore the term $\varepsilon P(x)$ when dealing with control issues in physical equipment. 
\end{remark}

\section{Experiments}\label{experiment}

\begin{figure}[t]
    \begin{minipage}[t]{0.48\columnwidth}
        \centering
        \includegraphics[width=0.8 \linewidth]{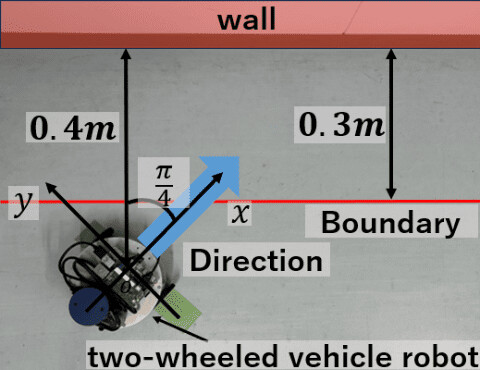}
        \caption{Environment of Exps.~1d and 1n.}
        \label{fig:experiment1_env}
    \end{minipage}
    \hspace{0.04\columnwidth}        
    \begin{minipage}[t]{0.48\columnwidth}
    \centering
        \includegraphics[width=0.8 \linewidth]{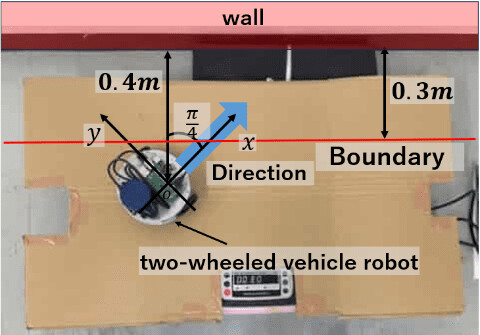}
    \caption{Environment of Exps.~2d and 2n.}
    \label{fig:experiment2_env}
    \end{minipage}
\end{figure}

\subsection{Parameter Settings}\label{parameters}
We describe the parameter values used in the experiments.

In the experimental environment, the number of the measurement points is $N=279$, the distance between the center of the sensor $S$ and the center of the driving wheels $B$ is $d=0.07$ m, and the distance between the center of the vehicle $O$ and $B$ is $e=0.025$ m. 

We design parameters as $\gamma=0.5$ and $\alpha=0.3$, and determine the diffusion coefficients as $c_1=0.035$ and $c_2=0$ using the estimation procedure shown in Appendix~\ref{yamaoka}. 
We also consider the deterministic controller $u=\phi_D$ for comparison, which is described in \eqref{eq:det-ctrl} with $K=\gamma=0.5$ and $C=0$, and derived by assuming $c_1=c_2=0$ in \eqref{cal_phiN}. 

Then, we set preinputs to $v_o=0.2$ and $w_o=0.2$, respectively, and the the initial state $x_{1 i^*}(0)= 0.4$ and $x_{2i^*}(0) =\pi/4$, where $i^*=i^*(t)$ is the subscript such that satisfies $x_{1i^*}-\alpha_{ci^*} = \min_{i \in \{1,2,\ldots,N\}} (x_{1i}-\alpha_{ci})$.

\subsection{Simulation and Experimental Results}
First, we set the vehicle on the floor as shown in Fig.~\ref{fig:experiment1_env} and perform numerical simulations and conduct experiments with the following conditions, respectively:
\begin{description}
    \item[\normalfont (Exp.~1d)]
    $u=\phi_D$;
    \item[\normalfont (Exp.~1n)]
    $u=\phi_N$ with $c_1=0.035$.
\end{description}

Then, we set the vehicle on the vibration platform as shown in Fig.~\ref{fig:experiment2_env}. and perform numerical simulations and conduct experiments with the following conditions, respectively:
\begin{description}
    \item[\normalfont (Exp.~2d)]
    $u=\phi_D$;
    \item[\normalfont (Exp.~2n)]
    $u=\phi_N$ with $c_1=0.035$.
\end{description}

\begin{figure}[t]
    \begin{minipage}[t]{0.48\columnwidth}
        \centering
        \includegraphics[width=0.8 \linewidth]{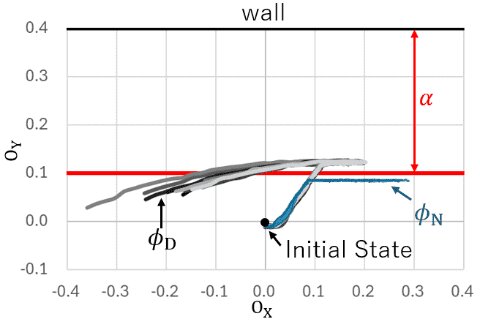}
        \caption{Trajectories of the center $O$ of the vehicle in Exps.~1d ($\phi_D$) and 1n ($\phi_N$).}
        \label{fig:trajectory}
    \end{minipage}
    \hspace{0.02\columnwidth}        
    \begin{minipage}[t]{0.50\columnwidth}
    \centering
        \includegraphics[width=0.8 \linewidth]{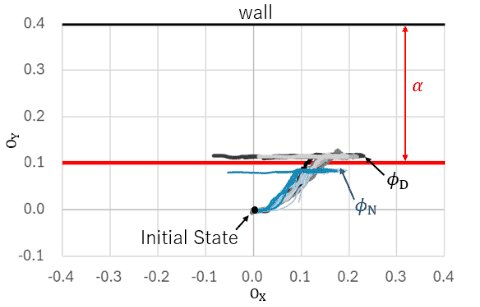}
    \caption{Trajectories of the center $O$ of the vehicle in Exps.~1d ($\phi_D$) and 1n ($\phi_N$).}
    \label{fig:trajectory_v}
    \end{minipage}
\end{figure}

The trajectories of Exps.~1d and 1n are shown in Fig.~\ref{fig:trajectory}, and Exps.~2d and 2n, Fig.~\ref{fig:trajectory_v}, respectively. The time responses of the states, the inputs, and the AS-RCBF $B(x)$ of Exps.~1d, 1n, 2d, and 2n are shown in Figs.~\ref{fig:rd_distance_d}--\ref{fig:rd_cbf_n} and Figs.~\ref{fig:rdv_distance_d}--\ref{fig:rdv_cbf_n}, respectively; the green and yellow lines show the time responses and the boundary $\alpha_{ci^*}$ of the numerical simulations, respectively. The blue and red lines also show the time responses and $\alpha_{ci^*}$ from ten trials of the experiments, respectively. 


\subsection{Discussion}\label{consideration}

In this subsection, we confirm the validity of the proposed controller via discussions on the results of the simulation and experimental results.

\begin{figure}[t]
    \begin{minipage}[t]{0.48\columnwidth}
    \centering
        \includegraphics[width=.9 \linewidth]{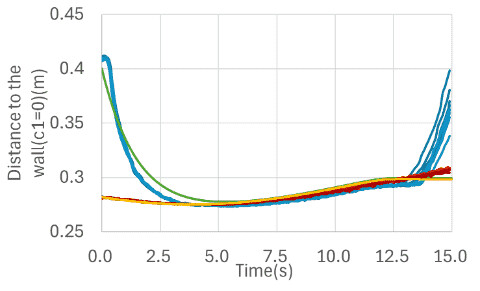}
        \caption{Time responses of $x_{1i^*}$ and $\alpha_{ci^*}$ in Exp.~1d. }
        \label{fig:rd_distance_d}

        \includegraphics[width=.9 \linewidth]{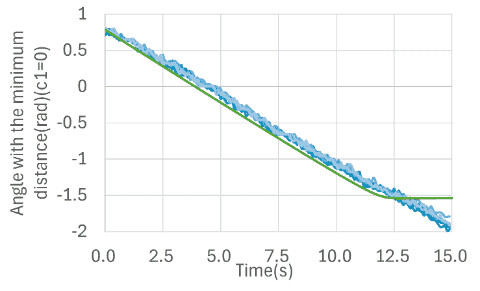}
        \caption{Time responses of $x_{2i^*}$ in Exp.~1d.}
        \label{fig:rd_angle_d}

        \includegraphics[width=.9 \linewidth]{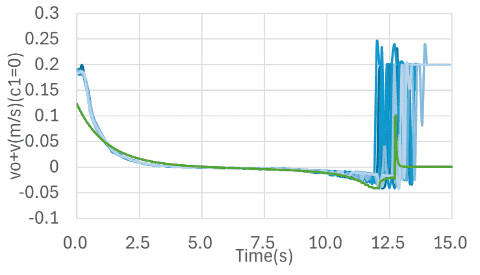}
        \caption{Time responses of $v_o + v$ in Exp.~1d.}
        \label{fig:rd_speed_d}
        
        \includegraphics[width=.9 \linewidth]{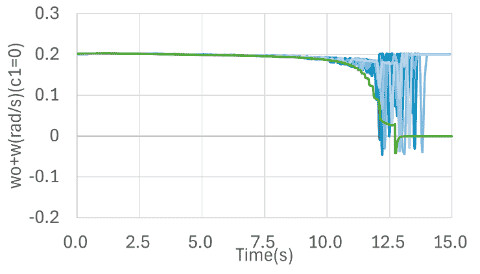}
        \caption{Time responses of $w_o + w$ in Exp.~1d.}
        \label{fig:rd_angspeed_d}

        \includegraphics[width=.9 \linewidth]{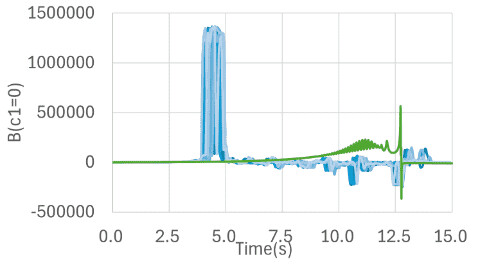}
        \caption{Time responses of $B(x)$ in Exp.~1d.}
        \label{fig:rd_cbf_d}
        \end{minipage}
    \hspace{0.04\columnwidth}        
    \begin{minipage}[t]{0.48\columnwidth}
    \centering
        \includegraphics[width=.9 \linewidth]{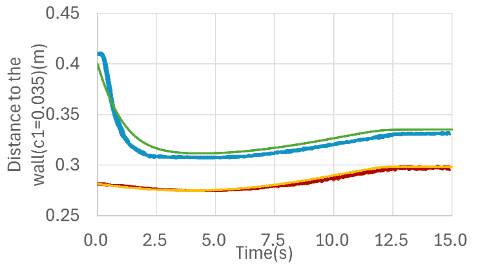}
        \caption{Time responses of $x_{1i^*}$ and $\alpha_{ci^*}$ in Exp.~1n. }
        \label{fig:rd_distance_n}

        \includegraphics[width=.9 \linewidth]{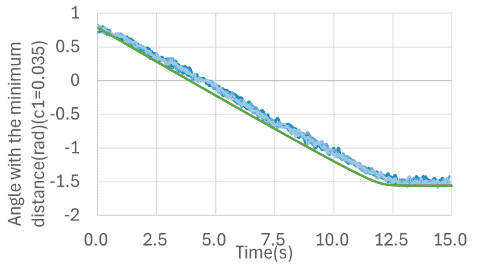}
        \caption{Time responses of $x_{2i^*}$ in Exp.~1n.}
        \label{fig:rd_angle_n}

        \includegraphics[width=.9 \linewidth]{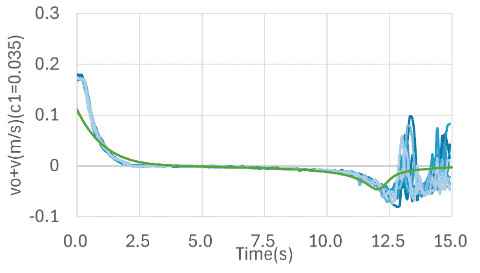}
        \caption{Time responses of $v_o + v$ in Exp.~1n.}
        \label{fig:rd_speed_n}
        
        \includegraphics[width=.9 \linewidth]{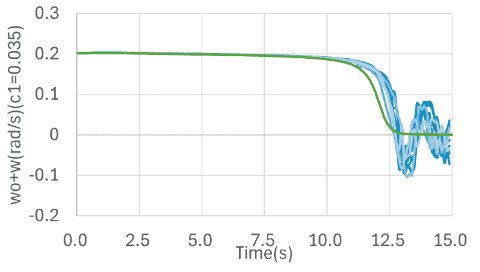}
        \caption{Time responses of $w_o + w$ in Exp.~1n.}
        \label{fig:rd_angspeed_n}

        \includegraphics[width=.9 \linewidth]{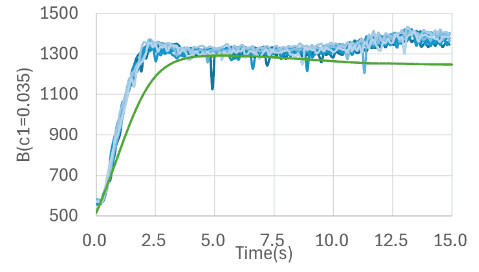}
        \caption{Time responses of $B(x)$ in Exp.~1n.}
        \label{fig:rd_cbf_n}
    \end{minipage}
\end{figure}

First, we compare the collision-avoiding results without adding noise. In Fig.~\ref{fig:rd_distance_d} (Exp.~1d), the deterministic safety-critical control $u=\phi_D$ is effective for avoiding collision to the threshold $\alpha_{ci^*}$ in simulation results, while $x_{1i^*}$ moved slightly past $\alpha_{ci^*}$ toward the wall. In contrast, in Fig.~\ref{fig:rd_distance_n} (Exp.~1n), the almost sure safety-critical control $u=\phi_N$ successfully avoided the collision. The difference is also shown in the trajectories shown in Fig.~\ref{fig:trajectory}. We consider that the cause of the failure in Exp.~1d is the existence of modeling and measurement errors, as well as variations in sample timing during actual measurements, and the cause of the success in Exp.~1n is that the safety compensation for noise is also effective for attenuating the errors and the variations.

\begin{figure}[t]
    \begin{minipage}[t]{0.48\columnwidth}
        \centering
        \includegraphics[width=.9 \linewidth]{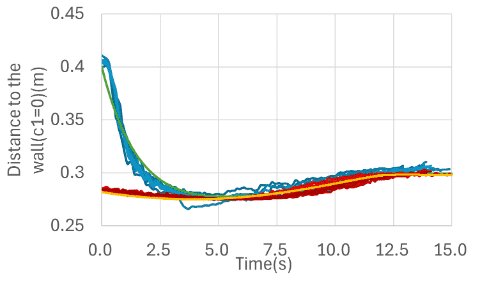}
        \caption{Time responses of $x_{1i^*}$ and $\alpha_{ci^*}$ in Exp.~2d. }
        \label{fig:rdv_distance_d}

        \includegraphics[width=.9 \linewidth]{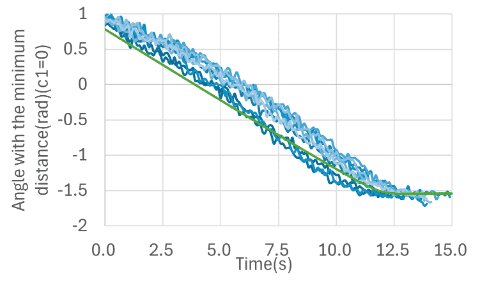}
        \caption{Time responses of $x_{2i^*}$ in Exp.~2d.}
        \label{fig:rdv_angle_d}

        \includegraphics[width=.9 \linewidth]{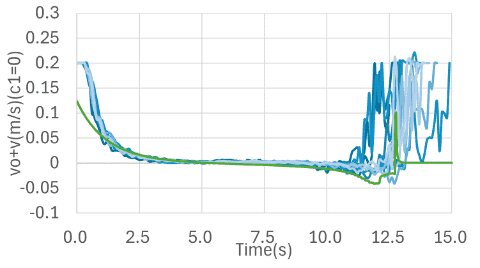}
        \caption{Time responses of $v_o + v$ in Exp.~2d.}
        \label{fig:rdv_speed_d}
        
        \includegraphics[width=.9 \linewidth]{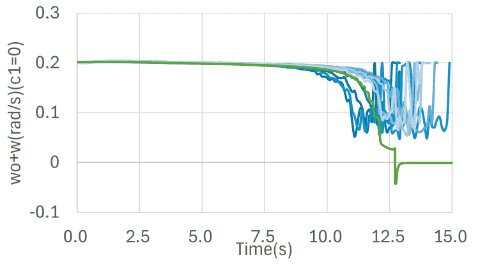}
        \caption{Time responses of $w_o + w$ in Exp.~2d.}
        \label{fig:rdv_angspeed_d}

        \includegraphics[width=.9 \linewidth]{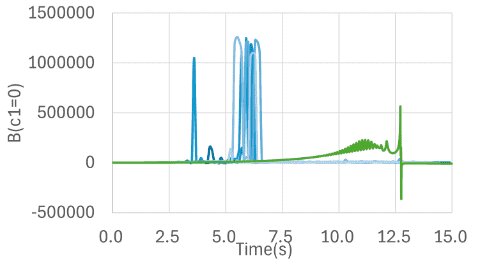}
        \caption{Time responses of $B(x)$ in Exp.~2d.}
        \label{fig:rdv_cbf_d}
    \end{minipage}
    \hspace{0.04\columnwidth}        
    \begin{minipage}[t]{0.48\columnwidth}
    \centering
        \includegraphics[width=.9 \linewidth]{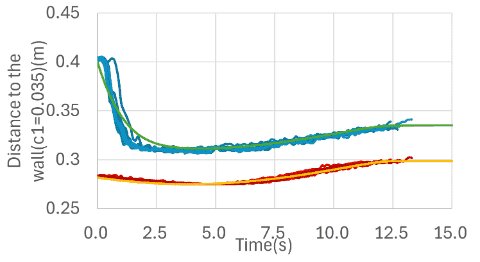}
        \caption{Time responses of $x_{1i^*}$ and $\alpha_{ci^*}$ in Exp.~2n. }
        \label{fig:rdv_distance_n}

        \includegraphics[width=.9 \linewidth]{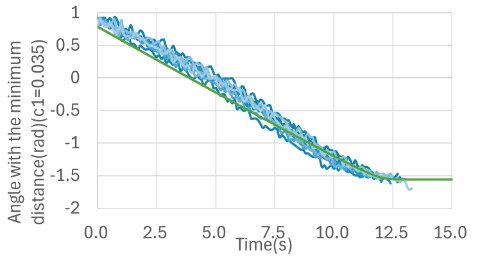}
        \caption{Time responses of $x_{2i^*}$ in Exp.~2n.}
        \label{fig:rdv_angle_n}

        \includegraphics[width=.9 \linewidth]{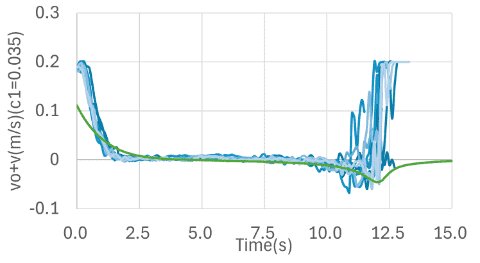}
        \caption{Time responses of $v_o + v$ in Exp.~2n.}
        \label{fig:rdv_speed_n}
        
        \includegraphics[width=.9 \linewidth]{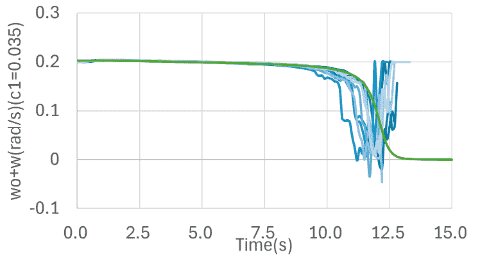}
        \caption{Time responses of $w_o + w$ in Exp.~2n.}
        \label{fig:rdv_angspeed_n}

        \includegraphics[width=.9 \linewidth]{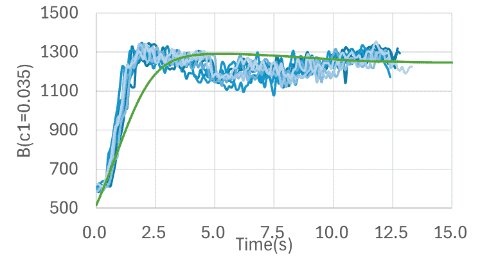}
        \caption{Time responses of $B(x)$ in Exp.~2n.}
        \label{fig:rdv_cbf_n}
        \end{minipage}
\end{figure}

Second, we compare the collision-avoiding results with adding noise. Comparing the trajectories in Fig.~\ref{fig:trajectory_v}, or, time responses of $x_{i^*}$ in Figs.~\ref{fig:rdv_distance_d} (Exp.~2d) and \ref{fig:rdv_distance_n} (Exp.~2n), the almost sure safety-critical control $u=\phi_N$ is effective for collision avoidance, while the deterministic safety-critical control $u=\phi_D$ fails the collision avoidance. 

Third, we compare the time responses of AS-RCBF $B(x)$. In Fig.~\ref{fig:rd_cbf_d} (Exp.~1d) and Fig.~\ref{fig:rdv_cbf_d} (Exp.~1n), the value of $B(x)$ rapidly becomes a large value as $B_{i^*}^{-1} =x_{1i^*}-\alpha_{ci^*}$, and thereafter it becomes negative. While $B(x)$ diverges as $B_{i^*}$ reaches zero theoretically, it is considered that the value remained finite due to a measurement error in actual measurement, and then, $B_{i^*}$ becomes negative in the subsequent time step. In contrast, in Fig.~\ref{fig:rd_cbf_n} (Exp.~1n) and Fig.\ref{fig:rdv_cbf_n} (Exp.~2n), $B(x)$ keeps positive finite value; it implies that the safety is ensured with 100\%. 

Finally, we consider rapid changes in input values. As in Figs.~\ref{fig:rd_speed_d}, \ref{fig:rd_angspeed_d}, \ref{fig:rd_speed_n}, \ref{fig:rd_angspeed_n}, \ref{fig:rdv_speed_d}, \ref{fig:rdv_angspeed_d}, \ref{fig:rdv_speed_n} and \ref{fig:rdv_angspeed_n}, the movements of $v_o + v$ and $w_o + w$ exhibit vibrations not caused by noise. The phenomena occur when the angle $x_{2i^*} \approx -\pi/2$ as shown in Figs.~\ref{fig:rd_angle_d}, \ref{fig:rd_angle_n}, \ref{fig:rdv_angle_d} and \ref{fig:rd_angle_n}; that is, the vehicle is facing parallel to the wall. Viewing \eqref{eq:lgb}, the signs of the elements of $L_{g}{B}$ both changes around $x_{2i^*}=-\pi/2$; this implies that the signs of $v_o+v$ and $w_o+w$ change frequently when $x_{2i^*} \approx -\pi/2$. Therefore, we estimate that the phenomena are due to the form of the proposed control law. To attenuate the movements, an appropriate design strategy for an AS-RCBF will be required. Otherwise, using a LiDAR with an extremely high number of measurement points, each element of $L_{g_i}B_i$ with a positive or negative sign is canceled out in the calculation of $L_gB$. The attenuation of the phenomena will be an important issue of future works.

\section{Conclusion}\label{conclusion}
In this paper, we designed an almost sure safety-critical control law based on an almost sure reciprocal control barrier function for stochastic systems proposed by Nishimura and Hoshino~\cite{nishimura_hoshino_RCBF}, using the state equation of the relative distance based on Kimura and Nishimoto~\cite{kimurakai}. We confirmed the validity of the designed control law that achieves collision avoidance via numerical simulations and we experiments. As a result, when the control law was designed without assuming noise, the vehicle crossed the boundary and safety was not maintained. In contrast, by incorporating noise into the control design, we achieved collision avoidance of the vehicle and maintained its safety. These results demonstrated that the almost sure safety-critical control law is effective even under stochastic vibration.

\section*{Acknowledgement(s)}

We would like to express our gratitude to Mr. Takumi Yamaoka at Kagoshima University and Mr. Kazuma Miyayoshi at Okayama University for their cooperation in conducting the experiments for this paper.

\appendix
\section{Estimation of Diffusion Coefficients}\label{yamaoka}
In this section, we estimate the values of the diffusion coefficients. For simplicity, we assume $c_2=0$ and estimate $c_1$ from preliminary vibration experiments on a vibration platform.

We set $x_{2i}=0$ and $v_o+v = 0$; that is,
\begin{align}
    dx_{1i} = c_1dw,
    \label{sigma_system}
\end{align}
and then, by applying the Euler-Maruyama scheme, we obtain the following discrete-time system
\begin{align}
    x_{1i}(t_{j+1})=x_{1i}(t_{j})+v(t_{j})\Delta t+c_1 \Delta w_j
    \label{risan_system},
\end{align}
where $j \in \{ 0,1,2,\ldots \} $, $\Delta t := t_{j+1}-t_j$ is constant for any $j$, and
\begin{align}
    c_1 \Delta w_j=x_{1i}(t_{j+1})-x_{1i}(t_{j})-v(t_j)\Delta t.
\end{align}
Thus, the variance of $c_1 \Delta w_j$ is calculated as 
\begin{align}
    \mathrm{Var}(c_1 \Delta w_j)
    &=\mathrm{Var}\left(x_{1i}(t_{j+1})-x_{1i}(t_{j})-v(t_j)\Delta t\right) \\
    &=c_1^2 \Delta t,
\end{align}
where $\mathrm{Var}(A)$ is the variance of $A$. 
Therefore, the noise coefficient $c_1$ is derived as
\begin{align}
    c_1
    =\sqrt{\frac{\mathrm{Var}\left(x_{1i}(t_{j+1})-x_{1i}(t_{j})-v(t_j)\Delta t\right)}{\Delta t}}
    \label{eq_sigma}.
\end{align}

Then, we identify the diffusion coefficient $c_1$ by conducting preliminary experiments. We place the vehicle on a vibration platform, actually vibrate it, and measure the noise ten times. The experiments result in $\mathrm{Var}(c_1 \Delta w_j)=0.00012$ and $\Delta t=0.1$, and then, the value of $c_1$ is identified as
\begin{align}
    c_1 = \sqrt{\frac{0.00012}{0.1}} \approx 0.035.
    \label{sigma=0.035}
\end{align}



\begin{thebibliography}{99}

\bibitem{takeuchi}
E.~Takeuchi. Development of mobile robots using middleware for robots, \emph{Journal of the Society of Instrument and Control Engineers} Vol.~57, No.~10, pp.~741--744, 2018.

\bibitem{maekawa}
K.~Maekawa. Autonomous mobile robot for factory automation and logistics automation, \emph{Systems, Control and Information} Vol.~64, No.~5, pp.~177--181, 2020.

\bibitem{artificial_potential_fuction}
E.~Rimon, D.E.~Koditschek. Exact robot navigation using artificial potential functions, \emph{IEEE Transactions on Robotics and Automation} Vol.~8, No.~5, pp.~501--518, 1992.

\bibitem{DWA}
D.~Fox, W.~Burgard and S.~Thrun. The dynamic window approach to collision avoidance, \emph{IEEE Robotics and Automation Magazine} Vol.~4, No.~1, pp.~23--33, 1997.

\bibitem{DWA2}
B.~Hahn. Enhancing obstacle avoidance in dynamic window approach via dynamic obstacle behavior prediction, \emph{Actuators} Vol.~14, No.~5, Article No.~207, 2025.

\bibitem{ND}
J.~Minguez, L.~Montano. Nearness diagram(ND) navigation: collision avoidance in troublesome scenarios, \emph{IEEE Transactions on Robotics and Automation} Vol.~20, No.~1, pp.~45--59, 2004.

\bibitem{planning_and_navigation}
T.~Tsubouchi, T.~Naniwa, S.~Arimoto. Planning and navigation by a mobile robot in the presence of multiple moving obstacles and their velocities, \emph{Journal of the Robotics Society of Japan} Vol.~12, No.~7, pp.~1029--1037, 1994.

\bibitem{fuzzy_inference}
Y.~Maeda, M.~Takegaki. Collision avoidance control among moving obstacles for a mobile robot on the fuzzy reasoning, \emph{Journal of the Robotics Society of Japan} Vol.~6, No.~6, pp.~518--522, 1988.

\bibitem{one_shot}
Z.~Zhang, G.~Hess, J.~Hu, E.~Dean, L.~Svensson, K.~\AA kesson. Future-oriented navigation: dynamic obstacle avoidance with one-shot energy-based multimodal motion prediction, \emph{IEEE Roborics and Automation Letters} Vol.~10, No.~8, pp.~8043--8050, 2025.

\bibitem{Ames_Coogan_}
A.D.~Ames, S.~Coogan, M.~Egerstedt, G.~Notomista, K.~Sreenath, P.~Tabuada. Control barrier functions: theory and applications, \emph{Proc.18th Euro. Control Conf.} , pp.~3420--3431, 2019.


\bibitem{kimurakai}
S.~Kimura, K.~Nishimoto. Collision avoidance human assist control with 2D LiDAR control barrier function, \emph{Transactions of the Society of Instrument and Control Engineers}, Vol.~61, No.~3, pp.~194--202, 2025.

\bibitem{kimuraiecon}
S.~Kimura. Control barrier function based on point cloud for human ssist control,
\emph{Proc. 46th Annual Conference of the IEEE Industrial Electronics Society (IECON)}, pp.~2645--2650, 2020.

\bibitem{nakamura}
H.~Nakamura, T.~Yoshinaga, Y.~Koyama, J.~Etoh. Control barrier function based human assist control, \emph{Transactions of the Society of Instrument and Control Engineers}, Vol.~55, No.~5, pp.~353--361, 2019.

\bibitem{nishimura_hoshino_RCBF}
Y.~Nishimura, K.~Hoshino. Control barrier functions for stochastic systems and safety-critical control designs, \emph{IEEE Transactions on Automatic Control}, Vol.~69, No.~11, pp.~1--8, 2024.

\bibitem{nishimura2018}
Y.~Nishimura and H.~Ito. Stochastic Lyapunov functions without differentiability at supposed equilibria, \emph{Automatica}, vol.~92, pp.~188--196, 2018.

\bibitem{henmi2026}
K.~Henmi, Y.~Nishimura, T.~Ikezaki and D.~Tabuchi. Almost sure front collision prevention control for an electric wheelchair via stochastic safety-critical control theory, \emph{Transactions of the Institute of Systems, Control and Information Engineers}, accepted in February 2026.

\end{thebibliography}
\end{document}